\documentclass[traditabstract]{aa} 
\usepackage{graphicx}
\usepackage{txfonts}
\usepackage{color}
\newcommand\kms{{\rm\,km\,s^{-1}}}
\newcommand\msun{\rm\,M_\odot}

\newcommand\hii{H\,{\sc ii} \,}

\begin{document}
\title{Search for OB stars running away from young star clusters.\\ II. The NGC\,6357 star-forming region}
\author{V.V.\ Gvaramadze\inst{1,2,3}
\and A.Y.\ Kniazev\inst{4,5} \and P.~Kroupa\inst{1} \and
S.~Oh\inst{1}}
\institute{Argelander-Institut f\"{u}r Astronomie, Universit\"{a}t
Bonn, Auf dem H\"{u}gel 71, 53121 Bonn, Germany \email{pavel;
skoh@astro.uni-bonn.de} \and Sternberg Astronomical Institute,
Moscow State University, Universitetskij Pr. 13, Moscow 119992,
Russia \email{vgvaram@mx.iki.rssi.ru} \and Isaac Newton Institute
of Chile, Moscow Branch, Universitetskij Pr. 13, Moscow 119992,
Russia \and South African Astronomical Observatory, PO Box 9, 7935
Observatory, Cape Town, South Africa \email{akniazev@saao.ac.za}
\and Southern African Large Telescope Foundation, PO Box 9, 7935
Observatory, Cape Town, South Africa}
\titlerunning{Bow shocks around young star clusters.\ II. NGC\,6357 region}
\authorrunning{Gvaramadze et al.}
\date{Received 22 July 2011/ Accepted 7 September 2011}
\abstract{Dynamical few-body encounters in the dense cores of
young massive star clusters are responsible for the loss of a
significant fraction of their massive stellar content. Some of the
escaping (runaway) stars move through the ambient medium
supersonically and can be revealed via detection of their bow
shocks (visible in the infrared, optical or radio). In this paper,
which is the second of a series of papers devoted to the search
for OB stars running away from young ($\la$ several Myr) Galactic
clusters and OB associations, we present the results of the search
for bow shocks around the star-forming region NGC\,6357. Using the
archival data of the {\it Midcourse Space Experiment} ({\it MSX})
satellite and the {\it Spitzer Space Telescope}, and the
preliminary data release of the Wide-Field Infrared Survey
Explorer (WISE), we discovered seven bow shocks, whose geometry is
consistent with the possibility that they are generated by stars
expelled from the young ($\sim 1$-2 Myr) star clusters, Pismis\,24
and AH03\,J1725-34.4, associated with NGC\,6357. Two of the seven
bow shocks are driven by the already known OB stars, HD\,319881
and [N78]\,34. Follow-up spectroscopy of three other
bow-shock-producing stars showed that they are massive (O-type)
stars as well, while the 2MASS photometry of the remaining two
stars suggests that they could be B0\,V stars, provided that both
are located at the same distance as NGC\,6357. Detection of
numerous massive stars ejected from the very young clusters is
consistent with the theoretical expectation that star clusters can
effectively lose massive stars at the very beginning of their
dynamical evolution (long before the second mechanism for
production of runaway stars, based on a supernova explosion in a
massive tight binary system, begins to operate) and lends strong
support to the idea that probably all field OB stars have been
dynamically ejected from their birth clusters. A by-product of our
search for bow shocks around NGC\,6357 is the detection of three
circular shells typical of luminous blue variable and late WN-type
Wolf-Rayet stars.}
\maketitle

\section{Introduction}
%
Close few-body dynamical encounters in the dense cores of young
massive star clusters are responsible for the loss of a
significant fraction of OB stars at the early stages of cluster
evolution (Poveda et al. \cite{po67}; Aarseth \& Hills
\cite{aa72}; Gies \cite{gi87}; Leonard \& Duncan \cite{le90};
Kroupa \cite{kr04}; Pflamm-Altenburg \& Kroupa \cite{pf06};
Moeckel \& Bate \cite{mo10}). The high central densities in young
clusters (the necessary condition for the production of runaway
stars) could be either primordial (e.g. Clarke \& Pringle
\cite{cl92}; Murray \& Lin \cite{mu96}; Clarke \& Bonnell
\cite{cl08}) or caused by dynamical mass segregation (e.g.
Portegies Zwart et al. \cite{po99}; G\"urkan et al. \cite{gu04};
Allison et al. \cite{al09}). In both cases, the clusters start to
eject stars long before the most massive cluster members explode
in supernovae, i.e. long before the binary-supernova ejection
mechanism (Blaauw \cite{bl61}) begins to operate. Moreover,
runaway stars could leave their parent clusters already during the
cluster formation process if a core of massive protostars forms
before bulk gas expulsion from the embedded cluster.

The runaway OB stars can be revealed either directly, via
measurement of their proper motions and/or radial velocities (e.g.
Moffat et al. \cite{mo98}; Mdzinarishvili \& Chargeishvili
\cite{md05}; Massey et al. \cite{ma05}; Evans et al. \cite{ev10};
Tetzlaff, Neuh\"{a}user \& Hohle \cite{te11}), or indirectly,
through the detection of bow shocks around them (Gvaramadze \&
Bomans \cite{gv08b}; Gvaramadze et al. \cite{gv10a}; Gvaramadze,
Kroupa \& Pflamm-Altenburg \cite{gv10d}; Gvaramadze,
Pflamm-Altenburg \& Kroupa \cite{gv11a}; Gvaramadze et al.
\cite{gv11b}). The latter possibility is especially helpful for
those runaways whose proper motions are still not available or are
measured with a low significance. The geometry of detected bow
shocks can be used to infer the direction of the stellar motion
and thereby to determine possible parent clusters for the
bow-shock-producing field stars (Gvaramadze \& Bomans
\cite{gv08b}; Gvaramadze et al. \cite{gv10a},d, \cite{gv11a}).

The present paper is the second of a series of papers devoted to
the search for OB stars running away from young ($\la$ several
Myr) Galactic clusters and OB associations. In Gvaramadze \&
Bomans (\cite{gv08b}; hereafter Paper\,I), we reported the
detection of three bow shocks produced by OB stars running away
from the star cluster \object{NGC\,6611}. Now we report the
discovery of seven bow shocks within $\sim 1\fdg5$ from the
star-forming region \object{NGC\,6357} and its associated
(massive) star clusters \object{Pismis\,24} and
\object{AH03\,J1725$-$34.4}. The geometry of all seven bow shocks
suggests that they are driven by stars expelled from NGC\,6357. In
Sect.\,\ref{sec:pismis-24} we review the relevant data for the
star clusters associated with NGC\,6357. Sect.\,\ref{sec:search}
presents the results of the search for bow shocks around
NGC\,6357. The bow-shock-producing stars are identified and
discussed in Sect.\,\ref{sec:stars}. In Sect.\,\ref{sec:shells} we
present the discovery of three circular shells around NGC\,6357.
Sect.\,\ref{sec:dis} deals with questions related to the content
of the paper. We summarize in Sect.\,\ref{sec:sum}.

\section{Star-forming region NGC\,6357 and its associated star
clusters}
\label{sec:pismis-24}

NGC\,6357 is an extended ($\sim 40\arcmin \times 60\arcmin$ or
$\sim 20$ pc $\times$ $30$ pc at a distance of 1.7 kpc; see below)
\hii region of ongoing massive star-formation in the Sagittarius
arm. At radio wavelengths NGC\,6357 is dominated by the two
components G\,353.2+0.9 and G\,353.1+0.6 (Schraml \& Mezger
\cite{sc69}; Felli et al. \cite{fe90}). Two open star clusters,
Pismis\,24 and AH03\,J1725$-$34.4, are associated with these
components, respectively. The first cluster was extensively
studied during the last years, while only sparse information
exists on the second one. Below we review the observational data
on both clusters in turn.

\subsection{Pismis\,24}

Pismis\,24 was recognized as an open star cluster by Pismis
(\cite{pi59}). It is believed that the massive stars of this
cluster are the main ionizing source responsible for the origin of
the \hii region NGC\,6357 (Lortet, Testor \& Niemela \cite{lo84};
Bohigas et al. \cite{bo04}; Cappa et al. \cite{ca11}). Line
observations of a molecular cloud associated with NGC\,6357 showed
that most of the molecular emission arises from the regions behind
or to the north of Pismis\,24, which indicates that the cluster is
immersed in a blister-like \hii region viewed face-on (Massi,
Brand \& Felli \cite{ma97}).

The cluster contains two very massive O3.5-type stars (Massey,
DeGioia-Eastwood \& Waterhouse \cite{ma01}; Walborn et al.
\cite{wa02}), one of which, Pismis\,24-1, was believed to be one
of the most massive known stars in the Galaxy, with an inferred
mass larger than $200 \, \msun$ (Walborn et al. \cite{wa02}).
Subsequent {\it Hubble Space Telescope} observations resolved
Pismis\,24-1 into two visual components, while follow-up
spectroscopy showed that one of the components
(\object{Pismis\,24-1SW}) is an O4\,III(f+) star (of mass of $\sim
100 \, \msun$) and the second one (\object{Pismis\,24-1NE}) is a
very massive ($\sim 100 \, \msun$) short-period spectroscopic
binary (Ma\'{i}z Apell\'{a}niz et al. \cite{ma07}).

Because of the high visual extinction towards Pismis\,24 ($\simeq
5-15$ mag; Wang et al. 2007), the stellar content of the cluster
was poorly studied until recently, so that only the most luminous
and less reddened cluster members (about two dozen altogether)
were identified, either by means of photometry or spectroscopy
(Moffat \& Vogt \cite{mo73}; Neckel \cite{ne78}, \cite{ne84};
Lortet et al. \cite{lo84}; Massey et al. \cite{ma01}). One of
these stars is a binary system (HD\,157504) composed of a WC7
Wolf-Rayet star (WR\,93) and an O7-9 star (van der Hucht
\cite{va01}).

Thanks to the ability of X-ray observations to penetrate heavy
absorption, the situation was improved with {\it Chandra}/ACIS
observations of NGC\,6357, which allowed the detection of even the
low-mass pre-main-sequence population of Pismis\,24 (with masses
extending down to $\sim 0.3 \, \msun$), thereby increasing the
number of the cluster members by a factor of $\sim 50$ and
bringing it to $\simeq 10000$ (Wang et al. \cite{wa07}). Most of
these stars are concentrated in a compact core of angular radius
of $\simeq 2$ arcmin (or $\simeq 1.0$ pc), which is centred on the
two most massive stars in the cluster, Pismis\,24-1 and
Pismis\,24-17\footnote{Note that the coordinates of Pismis\,24
given in the SIMBAD database are significantly ($\sim 16$ arcmin)
off the position of these stars and rather correspond to the
second star cluster, AH03\,J1725$-$34.4, associated with
NGC\,6357. The latter cluster is confused with Pismis\,24 in the
WEBDA database (http://www.univie.ac.at/webda/) as well.}, while
the remaining stars are spread with an approximately exponential
distribution over a halo of radius of $\simeq 11$ arcmin ($\simeq
5.4$ pc). The {\it Chandra} observation also detected two dozens
of X-ray sources, whose luminosities and colours suggest that they
could be OB stars. If spectroscopic follow-ups of these stars will
prove that they are indeed massive, then the OB star content of
Pismis\,24 will be doubled. Assuming that {\it Chandra} detected
all stars with mass $> 0.3 \, \msun$ and using the Kroupa
(\cite{kr01}) IMF (which is a two-part power-law IMF with a slope
$\alpha_{1} = 1.3$ for stellar masses between $0.08$ and $0.5
\msun$, and a Salpeter slope $\alpha_{2} = 2.3$ for more massive
stars), one finds a mass of the cluster of $\sim 10^4 \, \msun$
and the expected number of OB stars in the cluster of $\sim 100$.
The latter figure exceeds the number of confirmed and candidate OB
stars (i.e. 44 stars; Wang et al. \cite{wa07}) by a factor of 2,
from which one can conclude that either the IMF of the cluster's
members is steeper than the Salpeter one or most OB stars were
already ejected from the cluster (cf. Paper\,I; Pflamm-Altenburg
\& Kroupa \cite{pf06}; Moeckel \& Bate \cite{mo10})\footnote{The
discrepancy between the "observed" and expected number of OB stars
in Pismis\,24 would be less severe if most massive stars residing
in the cluster are binaries with mass ratio close to unity (cf.
Clarke \& Pringle \cite{cl92}).}. In the latter case, one can
expect to find numerous OB stars all around the cluster (see
Sect.\,\ref{sec:search}).

The simultaneous presence in Pismis\,24 of several very young
($\sim 1$ Myr) and very massive ($\simeq 90-100 \msun$) stars
(Ma\'{i}z Apell\'{a}niz et al. \cite{ma07}) and the more evolved
($\ga 2-2.5$ Myr old) WC7 star \object{WR\,93} (Massey et al.
\cite{ma01}) leads to an age discrepancy. To overcome this
discrepancy, one can assume that the star-formation process in the
cluster was not coeval or that WR\,93 is projected by chance near
the line of sight to the cluster (note that WR\,93 is located only
4 arcmin away from the cluster centre, i.e. well within the
cluster halo). In the latter case it is plausible that the parent
cluster of WR\,93 is associated with the same molecular cloud as
Pismis\,24 but is totally obscured by the dusty material of the
cloud, while the star itself is observable only because it is a
runaway, which already escaped from the densest part of the cloud
(cf. Gvaramadze et al. \cite{gv10a}). Note that the coexistence of
evolved and very young massive stars in the same star-forming
region is not a rare occurrence. For example, two evolved ($\sim
4$ Myr old) massive stars are projected against the young ($\simeq
1$ Myr) very massive cluster \object{NGC\,3603}, one of which, the
blue supergiant star \object{Sher\,25}, is separated from the
cluster core by less than 20 arcsec (or 1 pc in projection; e.g.
Melena et al. \cite{me08}; Rochau et al. \cite{ro10}). Another
possible explanation for the age discrepancy is that the most
massive stars in the cluster are blue stragglers, formed because
of merging of less massive stars in the course of close dynamical
encounters in the cluster's core (cf. Gvaramadze \& Bomans
\cite{08a}; Paper\,I).

Numerous distance estimates for Pismis\,24 range from $\simeq 1.7$
to 2.6 kpc (e.g. Moffat \& Vogt \cite{mo73}; Neckel \cite{ne78},
\cite{ne84}; Lortet et al. \cite{lo84}; Massey et al.
\cite{ma01}), putting the cluster in the Sagittarius spiral arm.
To constrain the distance to Pismis\,24, we used the optical
(Massey et al. \cite{ma01}) and near-infrared (2MASS; Skrutskie et
al. \cite{sk06}) photometry of all (six) known dwarf O stars in
the cluster (see Table\,2 in Massey et al. \cite{ma01}) and the
$UBVJHK$ synthetic photometry of Galactic O stars by Martins \&
Plez (\cite{ma06}). Assuming that the standard total-to-selective
absorption ratio $R_V =A_V /E(B-V) = 3.1$ is valid for Pismis\,24
and using the optical photometry, one finds a true distance
modulus of $\simeq 12.0$ mag and a distance of 2.5 kpc, which
agree well with those derived by Massey et al. (\cite{ma01}) on
the basis of a different calibration of stellar parameters. On the
other hand, adopting the extinction law from Rieke \& Rebofsky
(\cite{ri85}), so that
\begin{equation}
A_{K_{\rm s}} = 0.66E(J-K_{\rm s}) \, ,
\label{eqn:AK}
\end{equation}
and using the 2MASS photometry, one finds a true distance modulus
of $\simeq 11.15$ mag and a distance of 1.7 kpc. The discrepancy
between the two distance estimates could be understood if the
reddening towards Pismis\,24 is anomalous, i.e. $R_V > 3.1$. One
can show that $R_V \simeq 3.6$ would be enough to reconcile the
distance estimates (cf. with Neckel \& Chini \cite{ne81}, who
found $R_V =3.8$ for NGC\,6357). The distance of 1.7 kpc could
also be derived if one estimates the visual extinction $A_V$ using
the relationship
\begin{equation}
A_V \, = \, 1.39E(V-J) \, , \label{eqn:AV}
\end{equation}
which, according to Tapia et al. (\cite{ta88}), is valid across
the Galaxy, irrespective of high $R_V$ values (see also Persi \&
Tapia \cite{pe08}). This distance is consistent with that to the
H\,{\sc ii} region \object{NGC\,6334} ($\simeq 1.7$ kpc; Persi \&
Tapia \cite{pe08}), located on the sky at $\sim 2\degr$ to the
southwest from NGC\,6357. It is believed that both these H\,{\sc
ii} regions form a single high-mass star-forming complex (Russeil
et al. \cite{ru10} and references therein). In the following we
adopt a distance of 1.7 kpc for Pismis\,24 (and the star-forming
region NGC\,6357 as a whole) so that $1\degr$ corresponds to
$\simeq 30$ pc.

\subsection{AH03\,J1725$-$34.4}

The second cluster, AH03\,J1725$-$34.4, embedded in the
star-forming region NGC\,6357, is located at $\simeq 0\fdg26$ (or
$\simeq 7.5$ pc in projection) to the southeast from Pismis\,24
(Dias et al. \cite{di02}). The available information on this
cluster is very scarce. AH03\,J1725$-$34.4 contains at least four
OB stars (Neckel \cite{ne84}; Damke, Barba \& Morrell
\cite{da06}). Two of them, \object{[N78]\,49} and
\object{[N78]\,51}, are very massive ones, with spectral types
(O4\,III((f*)) and O3.5\,V, respectively; Damke et al.
\cite{da06}) comparable to those of the most massive stars in
Pismis\,24. Unfortunately, AH03\,J1725$-$34.4 was not covered by
the {\it Chandra} observations of NGC\,6357, so that the actual
extent and the stellar content of the cluster are still unknown.
The presence of two very massive stars in the cluster, however,
suggests that it should be as young ($\sim 1-2$ Myr) and at least
half as massive as Pismis\,24.

\section{Search for bow shocks}
\label{sec:search}

\begin{figure*}
\sidecaption \resizebox{14cm}{!}{\includegraphics{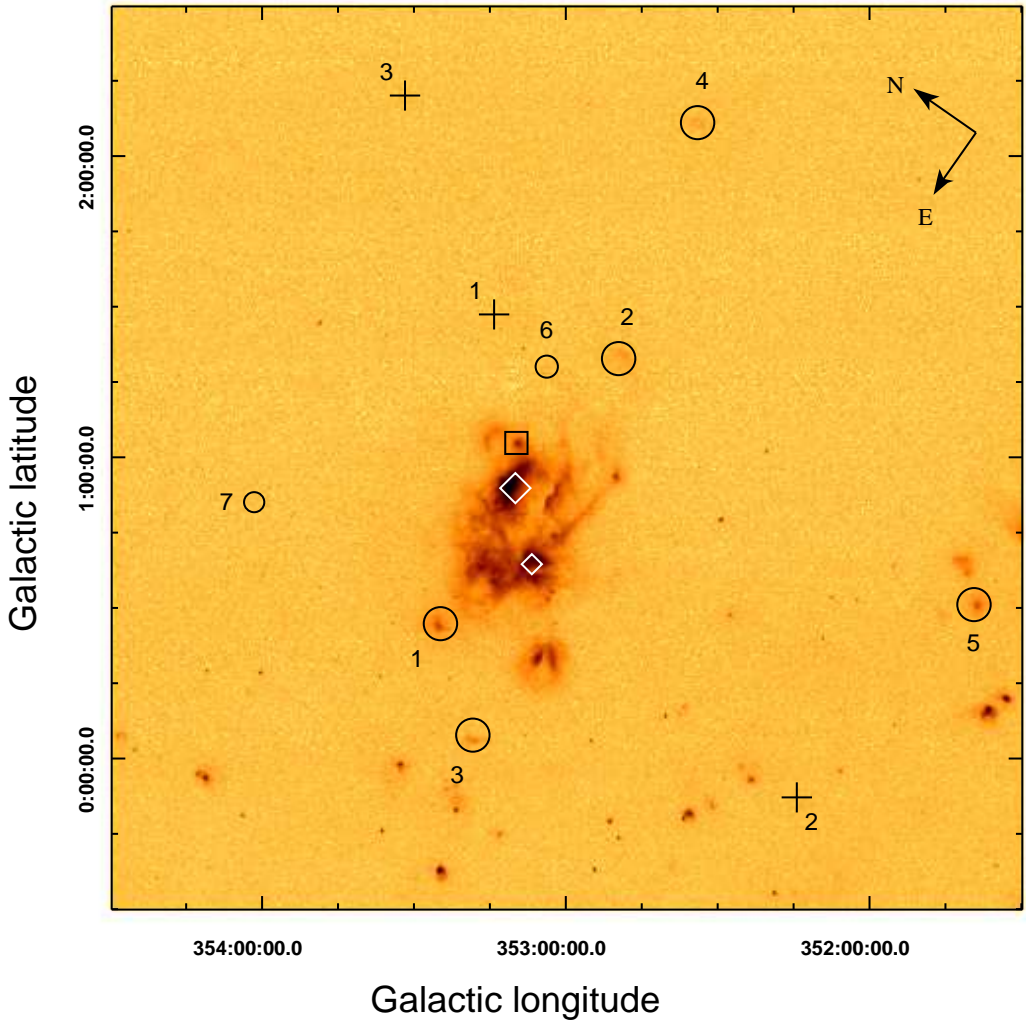}}
\caption{$3\degr \times 3\degr$ 21.3~$\mu$m ({\it MSX} band E)
image of the \hii region NGC\,6357 (containing the star clusters
Pismis\,24 and AH03\,J1725$-$34.4) and its environments. The
positions of seven bow shocks detected with {\it MSX} and {\it
Spitzer} are marked by large and small circles, respectively. The
position of the two most massive stars in Pismis\,24 (which define
the centre of the cluster) is indicated by a large diamond. The
position of the most massive star in the cluster
AH03\,J1725$-$34.4 is marked by a small diamond. The position of
IRAS\,17207$-$3404 is indicated by a square, while the positions
of three circular shells are indicated by crosses (see text for
details). }
  \label{fig:Pismis24}
\end{figure*}

A rich massive stellar content of Pismis\,24 and the presence of
very massive (binary) stars in both clusters associated with
NGC\,6357 suggest that this star-forming region was very efficient
in producing runaway stars (e.g. Gvaramadze \cite{gv07};
Gvaramadze, Gualandris \& Portegies Zwart \cite{gv09a}; Gvaramadze
\& Gualandris \cite{gv11}). One can, therefore, expect to find
some of the runaways via detection of their associated bow shocks
-- the natural attributes of supersonically moving stars (Baranov,
Krasnobaev \& Kulikovskii \cite{ba71}; Weaver et al. \cite{we77}).
The characteristic scale of a bow shock, $l$, depends on the
number density of the ambient medium, $n$, on the stellar
mass-loss rate, $\dot{M}$, and on the stellar space velocity,
$v_\ast$, as follows: $l\propto n^{-1/2}$, $l\propto \dot{M}
^{1/2}$ and $l\propto v_\ast ^{-1}$ (Baranov et al. \cite{ba71}).
The bow shocks are usually most prominent in the mid-infrared
(e.g. van Buren \& McCray \cite{va88}; van Buren, Noriega-Crespo
\& Dgani \cite{va95}; Paper\,I; Gvaramadze et al. \cite{gv10d},
\cite{gv11b}), but can also be detected in the optical (e.g. Kaper
et al. \cite{ka97}; Brown \& Bomans \cite{br05}; Paper\,I) and
radio (Benaglia et al. \cite{be10}) wavebands. It should be noted
that only a minority ($\simeq 20$ per cent) of runaway OB stars
are associated with (detectable) bow shocks (van Buren
\cite{va93}; van Buren et al. \cite{va95}; Huthoff \& Kaper
\cite{hu02}; Gvaramadze et al. \cite{gv10d}). The paucity of the
bow-shock-producing stars is mostly due to the fact that the
majority of runaway stars are moving through a low-density, hot
medium, so that the emission measure of their bow shocks is below
the detection limit or the bow shocks cannot be formed at all
because the sound speed in the local interstellar medium is higher
than the stellar space velocity (Kaper, Comer\'{o}n \& Barziv
\cite{ka99}; Huthoff \& Kaper \cite{hu02}). Moreover, the bow
shocks generated by runaway stars receding from NGC\,6357 and
interacting with the dense material of the background (parent)
molecular cloud (see Sect.\,\ref{sec:pismis-24}) would be very
compact and therefore hardly detectable, while those projected
against the \hii region might be hidden by its bright emission
(see Fig.\,\ref{fig:Pismis24}). From this it follows that the
actual number of stars ejected from the clusters in NGC\,6357
could several times exceed the number of detected
bow-shock-producing stars.

To search for bow shocks around NGC\,6357, we used the archival
data from the Mid-Infrared Galactic Plane Survey (Price et al.
2001), the 24 and 70 Micron Survey of the Inner Galactic Disk with
MIPS\footnote{MIPS = Multiband Imaging Photometer for {\it
Spitzer} (Rieke et al. \cite{ri04}).} (MIPSGAL; Carey et al.
\cite{ca09}) and the preliminary data release of the Mid-Infrared
All Sky Survey carried out with the Wide-field Infrared Survey
Explorer (WISE; Wright et al. 2010). The first survey, carried out
with the Spatial Infrared Imaging Telescope onboard the {\it
Midcourse Space Experiment} ({\it MSX}) satellite, covers the
entire Galactic plane within $|b|<5\degr$ and provides images at
18 arcsec resolution in four mid-infrared spectral bands centred
at 8.3~$\mu$m (band A), 12.1~$\mu$m (band C), 14.7~$\mu$m (band
D), and 21.3~$\mu$m (band E). The MIPSGAL survey (carried out with
the {\it Spitzer Space Telescope}) mapped 278 square degrees of
the inner Galactic plane ($|b|<1\degr$ is covered for $5\degr < l
< 63\degr$ and $298\degr < l <355\degr$ and $|b|<3\degr$ is
covered for $|l|<5\degr$) and provides 24\,$\mu$m images at 6
arcsec resolution. The current release of the WISE survey covers
57 per cent of the sky and provides images at four wavelengths:
3.4, 4.6, 12 and 22\,$\mu$m, with angular resolution of 6.1, 6.4,
6.5 and 12.0 arcsec, respectively. The advantage of the {\it MSX}
and WISE surveys is that they cover the whole Galactic plane and
extend to higher Galactic latitudes than the MIPSGAL one, thereby
allowing us to search for high-velocity runaways ejected at large
angles to the Galactic plane (which could be off the region
covered by the MIPSGAL survey). On the other hand, the better
angular resolution of the MIPSGAL survey could be vital for the
detection of bow shocks generated by stars moving through the
dense gas of the parent molecular cloud. To search for possible
optical counterparts to the bow shocks and to identify their
associated stars, we used the Digitized Sky Survey II (DSS-II;
McLean et al. \cite{mc00}).

Assuming that the age of the star clusters in NGC\,6357 is $\sim
1-2$ Myr and adopting the typical velocity of runaway stars of
$\sim 30-50 \, \kms$, one finds that stars leaving the clusters at
the very beginning of their dynamical evolution would be confined
within $\sim 1-3\degr$ from NGC\,6357. Thus, one can neglect the
effect of the Galactic gravitational potential on their
trajectories. Correspondingly, one can expect that the bow shocks
produced by the ejected stars would be directed away from their
parent clusters, provided that there are no peculiar large-scale
flows in the interstellar medium through which the stars are
moving.

\begin{figure}
\resizebox{9cm}{!}{\includegraphics{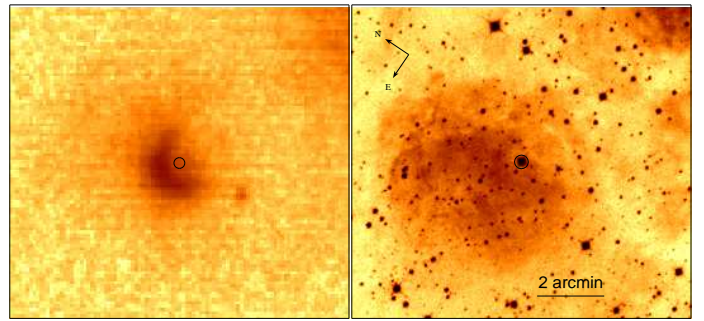}}
 \caption{{\it Left:} {\it MSX} 21.3~$\mu$m image of bow shock\,1.
 {\it Right:} DSS-II (red band) image of the same field. The position of
 the bow-shock-producing star (star\,1) is marked by a circle.
 }
  \label{fig:bow1}
\end{figure}

\begin{figure}
 \resizebox{9cm}{!}{\includegraphics{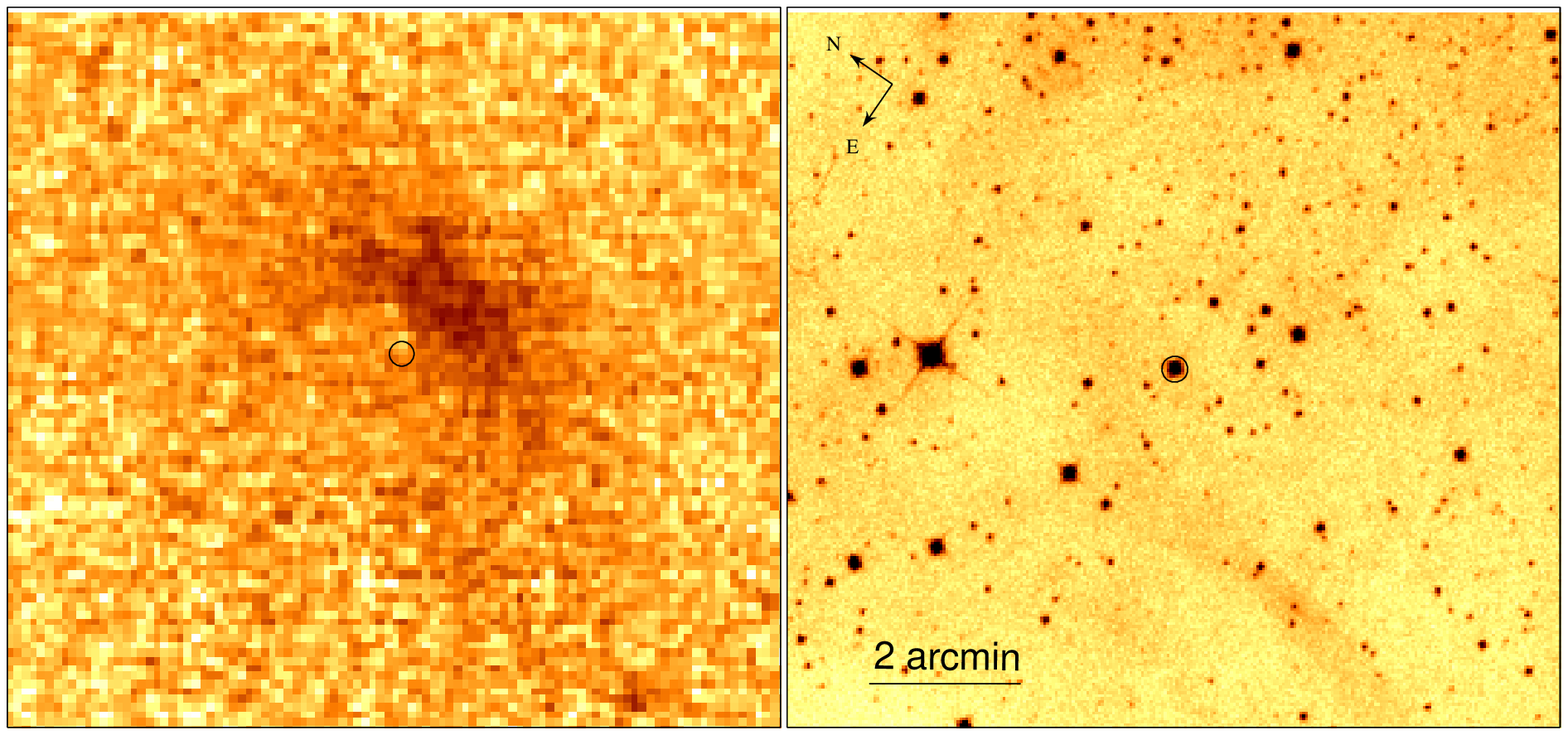}}
 \caption{{\it Left:} {\it MSX} 21.3~$\mu$m image of bow shock\,2.
 {\it Right:} DSS-II (red band) image of the same field. The
position of the bow-shock-producing star (star\,2) is marked by a
circle.}
  \label{fig:bow2}
\end{figure}

\begin{figure}
 \resizebox{9cm}{!}{\includegraphics{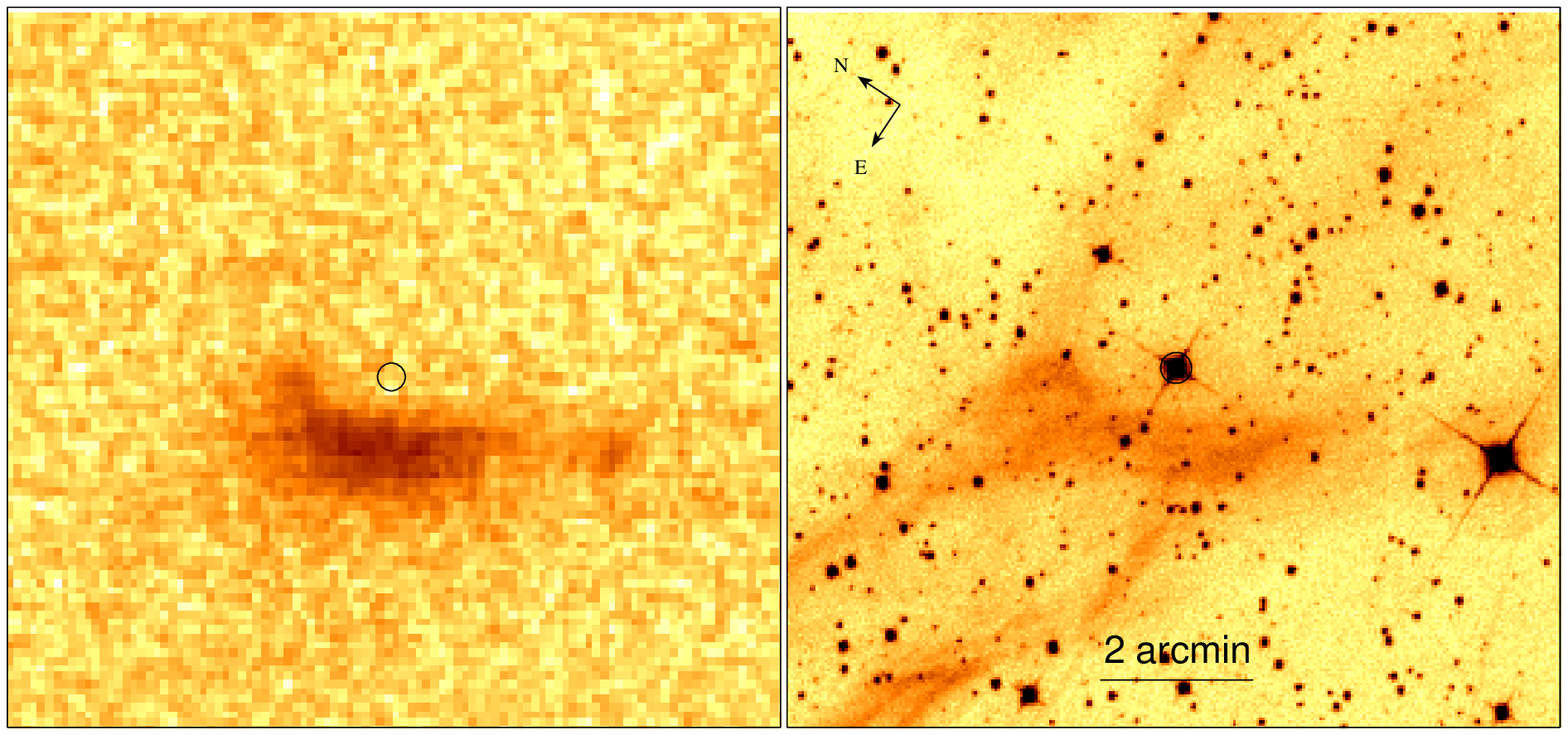}}
 \caption{{\it Left:} {\it MSX} 21.3~$\mu$m image of bow shock\,3.
 {\it Right:} DSS-II (red band) image of the same field.
The position of the bow-shock-producing star (star\,3 or
HD\,319881) is marked by a circle.}
  \label{fig:bow3}
\end{figure}

\begin{figure}
 \resizebox{9cm}{!}{\includegraphics{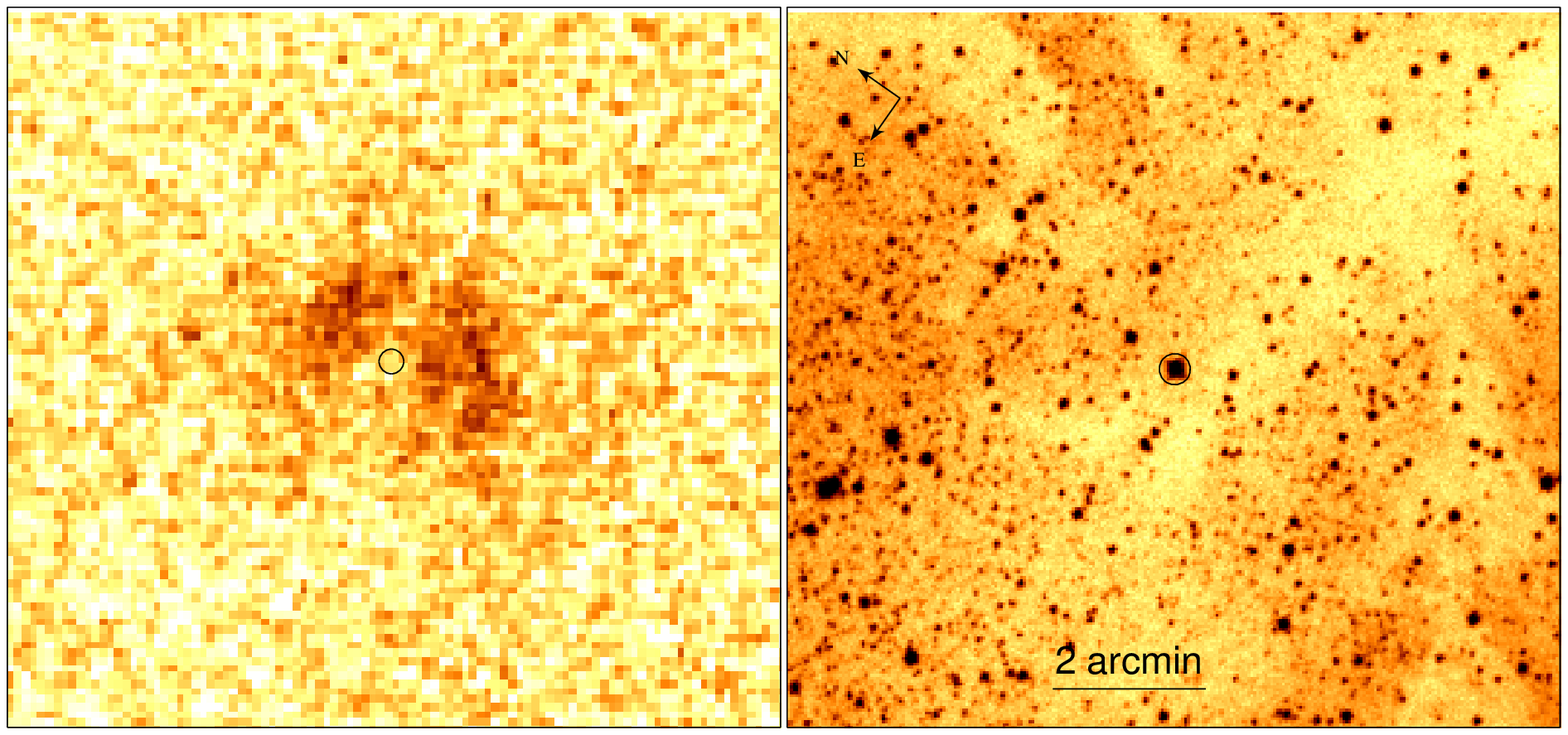}}
 \caption{{\it Left:} {\it MSX} 8.3~$\mu$m image of bow shock\,4.
 {\it Right:} DSS-II (red band) image of the same field. The
position of the bow-shock-producing star (star\,4) is marked by a
circle.}
  \label{fig:bow4}
\end{figure}

\begin{figure}
 \resizebox{9cm}{!}{\includegraphics{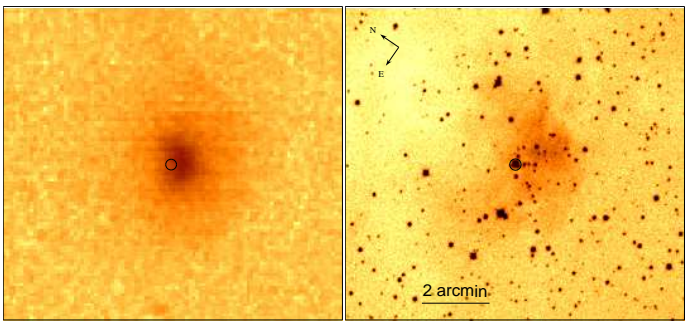}}
 \caption{{\it Left:} {\it MSX} 21.3~$\mu$m image of bow shock\,5.
 {\it Right:} DSS-II (red band) image of the same field. The
position of the bow-shock-producing star (star\,5 or [N78]\,34) is
marked by a circle.}
  \label{fig:bow5}
\end{figure}

First, we searched for bow shocks using the {\it MSX} data. The
search was carried out in a $12\degr$ wide area elongated along
the Galactic plane and centred at the longitude of NGC\,6357 ($l
\simeq 353\degr$). Along the Galactic latitude, the search was
limited by the {\it MSX} coverage ($|b|<5\degr$), so that
potentially we were able to detect bow shocks produced by stars
leaving the cluster immediately after its formation and moving
perpendicularly to the Galactic plane with a peculiar (tangential)
velocity of $\sim 100 \, \kms$. The visual inspection of {\it MSX}
21.3~$\mu$m images revealed five bow shocks (indicated in
Fig.\,\ref{fig:Pismis24} by large circles). All of them have a
clear arc-like structure that opens towards NGC\,6357
(Figs.\,\ref{fig:bow1}-\ref{fig:bow5}), which suggests that these
structures are generated by stars expelled from this star-forming
region. Bow shock\,4 has a prominent gap on its leading edge
(Fig.\,\ref{fig:bow4}; see also below). All five bow shocks are
most prominent at 21.3~$\mu$m, although some of them are visible
in other {\it MSX} wavebands as well. Bow shock 3 has an obvious
optical counterpart in the DSS-II (see Fig.\,\ref{fig:bow3}),
while bow shocks 1 and 5 are embedded in \hii regions (the \hii
region associated with bow shock 5 is known as \object{GAL
351.66+00.52}; Lockman \cite{lo89}).


\begin{figure}
 \resizebox{9cm}{!}{\includegraphics{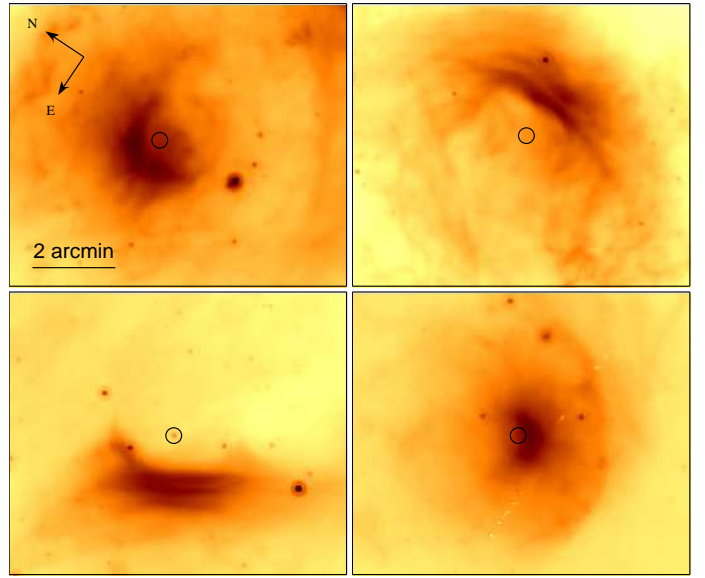}}
 \caption{From left to right, and from top to bottom:
MIPS $24\,\mu$m images of bow shocks\,1, 2, 3, and 5. The
positions of the associated stars are marked by circles. The
orientation and the scale of the images are the same.}
  \label{fig:MIPSbows}
\end{figure}

\begin{figure}
 \resizebox{9cm}{!}{\includegraphics{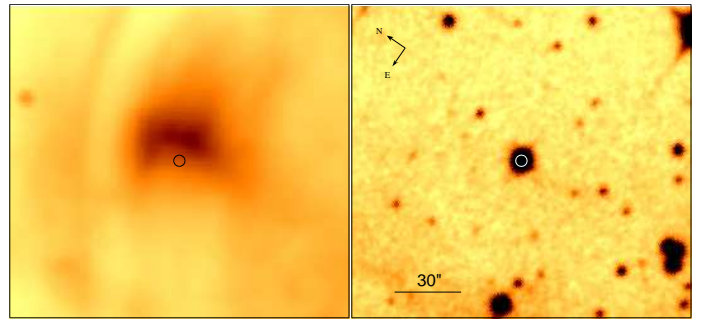}}
 \caption{{\it Left:} MIPS $24\,\mu$m image of bow shock\,6.
 {\it Right:} DSS-II (red band) image of the same field. The
position of the bow-shock-producing star (star\,6) is marked by a
circle.}
  \label{fig:bow6}
\end{figure}

\begin{figure}
 \resizebox{9cm}{!}{\includegraphics{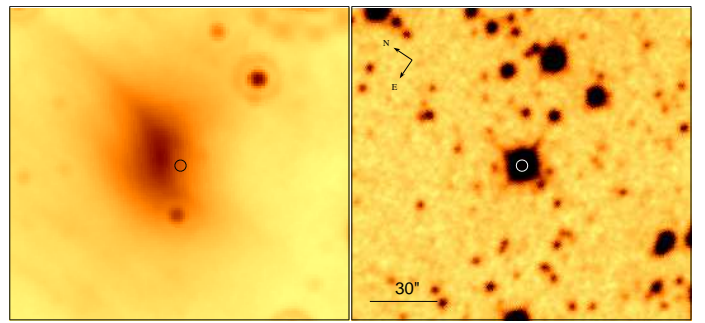}}
 \caption{{\it Left:} MIPS $24\,\mu$m image of bow shock\,7.
 {\it Right:} DSS-II (red band) image of the same field. The
position of the bow-shock-producing star (star\,7) is marked by a
circle.}
  \label{fig:bow7}
\end{figure}

Then, we used the data from the MIPSGAL survey. Four of the bow
shocks discovered with {\it MSX} were covered by this survey and
we give their MIPS 24\,$\mu$m images in Fig.\,\ref{fig:MIPSbows}.
We also discovered two new bow shocks whose orientation is
consistent with the possibility that their associated stars were
ejected from NGC\,6357 (see Figs.\,\ref{fig:bow6} and
\ref{fig:bow7} and compare them with Fig.\,\ref{fig:Pismis24},
where the positions of these bow shocks are indicated by small
circles). The angular extent of both bow shocks ($\sim 30$ arcsec)
is several times smaller than that of the bow shocks discovered
with {\it MSX}, which can be understood if the stars generating
these two bow shocks are either moving through the denser ambient
medium (e.g. the background molecular cloud) or/and have weaker
winds and higher peculiar velocities. Neither of these bow shocks
were detected in the optical range.


\begin{figure}
 \resizebox{9cm}{!}{\includegraphics{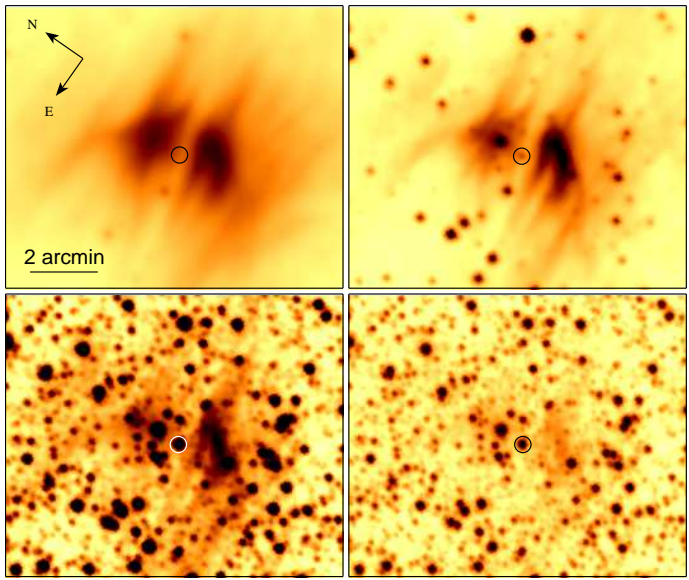}}
 \caption{From left to right, and from top to bottom:
WISE 22, 12, 4.6 and 3.4~$\mu$m images of bow shock\,4. The
position of the associated star (star\,4) is marked by a circle.
The orientation and the scale of the images are the same.}
  \label{fig:WISEbow}
\end{figure}

The use of the WISE data did not result in discovery of new bow
shocks around NGC\,6357. All seven bow shocks detected with {\it
MSX} and {\it Spitzer} are clearly visible in the WISE data as
well. In Fig.\,\ref{fig:WISEbow} we present images of bow shock\,4
(not covered by the MIPSGAL survey) at all four WISE wavelenghts,
showing its curious fine structure (most prominent at 12 and
22~$\mu$m). The possible origin of the gap in the leading edge of
bow shock\,4 and the cirrus-like filaments around it are discussed
in Sect.\,\ref{sec:dis}.

\begin{figure}
 \resizebox{9cm}{!}{\includegraphics{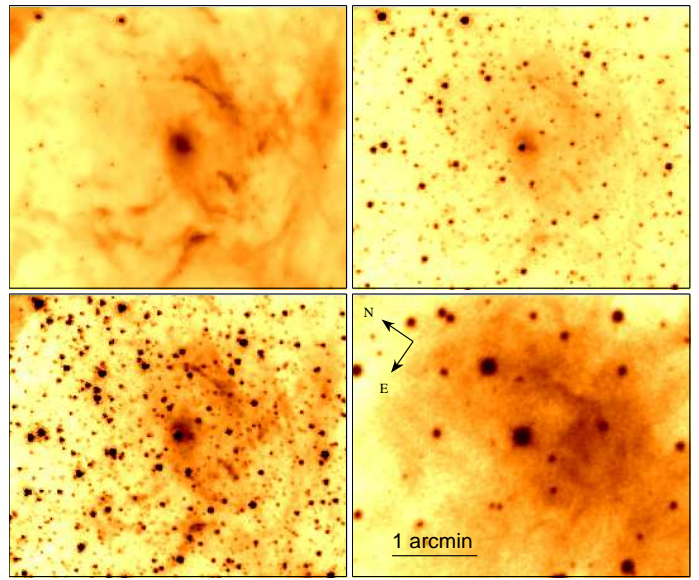}}
 \caption{From left to right, and from top to bottom:
IRAC 8, 4.5 and 3.6 $\mu$m, and DSS-II (red band) images of the
candidate bow-shock-producing star (star\,8) embedded in the
extended region of infrared emission (known as
IRAS\,17207$-$3404). The orientation and the scale of the images
are the same. }
  \label{fig:nebula}
\end{figure}

A by-product of our search for bow shocks using the {\it Spitzer}
data is the discovery of a compact arcuate nebula attached to one
of the candidate OB stars revealed with {\it Chandra} (namely,
star\,9 from Table\,7 of Wang et al. 2007). The MIPS 24\,$\mu$m
image of this nebula is saturated, so that we give in
Fig.\,\ref{fig:nebula} the {\it Spitzer} 8, 4.5 and $3.6\,\mu$m
images obtained with the Infrared Array Camera (IRAC; Fazio et al.
\cite{fa04}) within the Galactic Legacy Infrared Mid-Plane Survey
Extraordinaire (GLIMPSE; Benjamin et al. \cite{be03}). The
orientation and the small angular size of the nebula (embedded in
the more extended region of infrared emission known as
IRAS\,17207$-$3404; marked in Fig\,\ref{fig:Pismis24} by a square)
are consistent with the interpretation of the nebula as a bow
shock generated by a massive star ejected from Pismis\,24 and
moving through the dense background molecular cloud.
Correspondingly, the extended {\it IRAS} source can be interpreted
as an \hii region created by the ionizing emission of the star
(cf. Paper\,I; Gvaramadze et al. \cite{gv11b}). Alternatively, the
nebula could be a circumstellar (toroidal) shell, similar to those
observed around some evolved massive stars (see, e.g., Fig.\,4 in
Gvaramadze, Kniazev \& Fabrika \cite{gv10c} for the IRAC
5.8\,$\mu$m image of a nebula around the candidate luminous blue
variable star MN\,56). The 2MASS photometry of the star, however,
suggests that it could be a main-sequence late O-type star (see
Sect.\,\ref{sec:stars}), so that we consider the bow-shock
interpretation of the arcuate nebula as more preferable. Note that
the candidate OB star associated with the nebula is separated from
the centre of Pismis\,24 by only $\simeq 9$ arcmin, i.e. it is
located (at least in projection) well within the cluster halo. For
the sake of completeness, we give the details of this star
(hereafter star\,8) in Table\,\ref{tab:phot}.

Another by-product of the search for bow shocks with {\it Spitzer}
is the discovery of three ring-like shells with central point
sources. We present these shells and discuss their origin in
Sect.\,\ref{sec:shells}.

\section{Bow-shock-producing stars}
\label{sec:stars}

Inspection of the DSS-II (red band) images suggested that the bow
shocks discovered with {\it MSX} and {\it Spitzer} are associated
with stars indicated in Figs.\,\ref{fig:bow1} -- \ref{fig:bow7} by
circles. The details of these stars are given in
Table\,{\ref{tab:phot}. Columns 2, 3, 5, and 6 give the equatorial
coordinates of the stars and their $J$- and $K_{\rm S}$-magnitudes
(all taken from the 2MASS catalogue; Skrutskie et al.
\cite{sk06}), while column 4 gives the visual magnitudes of the
stars taken from the NOMAD catalogue (Zacharias et al. 2004).
Columns 7 and 8 give the $K_{\rm S}$-band interstellar extinction
towards the stars and their absolute magnitudes in the same band
derived under the assumption that the stars are located at the
same distance as NGC\,6357. Column 9 provides spectroscopic and
photometric (enclosed in brackets) spectral types of the stars
(see below).

\begin{table*}
  \caption{Details of seven bow-shock-producing stars (stars\,1--7) and one candidate bow-shock-producing star
  (star\,8) detected around NGC\,6357. The last column provides spectroscopic and photometric
(enclosed in brackets) spectral types of the stars (see text for
details).}
  \label{tab:phot}
  \renewcommand{\footnoterule}{}
\begin{tabular}{lcccccccccc}
\hline \hline
Star & RA (2000) & Dec. (2000) & $V$ & $J$ & $K_{\rm s}$ & $A_{K_{\rm s}}$ & $M_{K_{\rm s}}$ & Spectral type  \\
\hline
1 & 17 27 11.23 & $-$34 14 34.9 & 11.77 & 8.42 & 7.84 & 0.52 & $-3.84$ & O7.5:$^{a}$ (O7\,V) \\
2 & 17 22 03.43 & $-$34 14 24.1 & 13.59 & 8.94 & 8.01 & 0.75 & $-3.89$ & O5.5:$^{a}$ (O6.5-7\,V) \\
3 or HD\,319881 & 17 28 21.67 & $-$34 32 30.3 & 10.14 & 7.54 & 7.10 & 0.43 & $-4.48$ & O6\,Vn$^{b}$ (O5\,V) \\
4 & 17 18 15.40 & $-$34 00 06.1 & 11.80 & 8.19 & 7.53 & 0.57 & $-4.19$ & O6.5:$^{a}$ (O6\,V) \\
5 or [N78]\,34 & 17 22 05.62 & $-$35 39 55.5 & 13.11 & 8.68 & 7.82 & 0.71 & $-4.04$ & O8/B0\,III/V$^{c}$ (O6.5\,V) \\
6 & 17 22 50.02 & $-$34 03 22.4 & 13.90 & 9.68 & 8.81 & 0.71 & $-3.05$ & (B0\,V) \\
7 & 17 27 12.53 & $-$33 30 40.0 & 12.18 & 9.21 & 8.65 & 0.51 & $-3.01$ & (B0\,V) \\
\hline
8 & 17 24 05.62 & $-$34 07 09.5 & 11.74 & 9.03 & 8.46 & 0.51 & $-3.20$ & (O9-9.5\,V) \\
\hline
\end{tabular}
 \tablefoot{
 \tablefoottext{a}{this work.}
 \tablefoottext{b}{Drilling \& Perry (\cite{dr81}).}
 \tablefoottext{c}{Neckel (\cite{ne84}).}
 }
\end{table*}

Using the SIMBAD
database\footnote{http://simbad.u-strasbg.fr/simbad/} and the
VizieR catalogue access
tool\footnote{http://webviz.u-strasbg.fr/viz-bin/VizieR}, we found
that two of the five bow shocks discovered with {\it MSX} are
generated by known OB stars (with spectroscopically derived
spectral types). Namely, it turns out that bow shocks\,3 and 5 are
produced by the O6\,Vn (Drilling \& Perry \cite{dr81}) star
\object{HD\,319881} and the O8/B0\,III/V (Neckel \cite{ne84}) star
\object{[N78]\,34}, respectively. To determine the nature of the
other three stars (hereafter stars\,1, 2 and 4), we obtained their
spectra with the South African Astronomical Observatory (SAAO)
1.9-m telescope (see Sect.\,\ref{sec:spec}). The stars associated
with the bow shocks discovered with {\it Spitzer} (hereafter
stars\,6 and 7) still have to be observed spectroscopically. If,
however, we assume that these two stars are located at the same
distance as NGC\,6357, then one can estimate their spectral types
using the 2MASS photometry and calibration of absolute magnitudes
and intrinsic colours (see Sect.\,\ref{sec:spect}).

\subsection{Spectral types of the bow-shock-producing stars\,1, 2 and 4}
\label{sec:spec}

The observations of stars\,1, 2 and 4 were performed with the
Cassegrain spectrograph using a long slit of 3$^{\prime}
\times$1.8$^{\prime\prime}$ on 2009 June 1 and 2. Grating with 300
lines mm$^{-1}$ was used with the 266$\times$1798 pixels CCD
detector SITe with two setups with the wavelength coverage of
$\sim 3760-7750$ \AA\ for stars\,1 and 4 and $\sim 3560-7550$ \AA\
for star 2. For both setups the spectral resolution was $\sim$2.3
\AA~pixel$^{-1}$ (FWHM $\sim$7~\AA) along the dispersion direction
and the scale along the slit was 1\farcs4 pixel$^{-1}$. The seeing
during the observations varied from 1.0 to 2.1 arcsec from object
to object, but was stable for each object. Exposure times were
$3\times 300$, $3\times 600$ and $3\times 300$ sec for stars\,1, 2
and 4, respectively. The observations of star 2 were made in
variable conditions, with drastic drops in transparency, resulting
in the worse quality of the resulting spectrum. Spectra of Cu--Ar
comparison arcs were obtained to calibrate the wavelength scale.
The spectrophotometric standard stars LTT\,7379 and LTT\,3864 were
observed with a slit width of 5 arcsec at the beginning and at the
end of the nights for flux calibration. Data reduction was
performed using standard procedures (see, e.g., Kniazev et al.
\cite{kn08} for details).

\begin{figure*}
\includegraphics[width=12cm,angle=270]{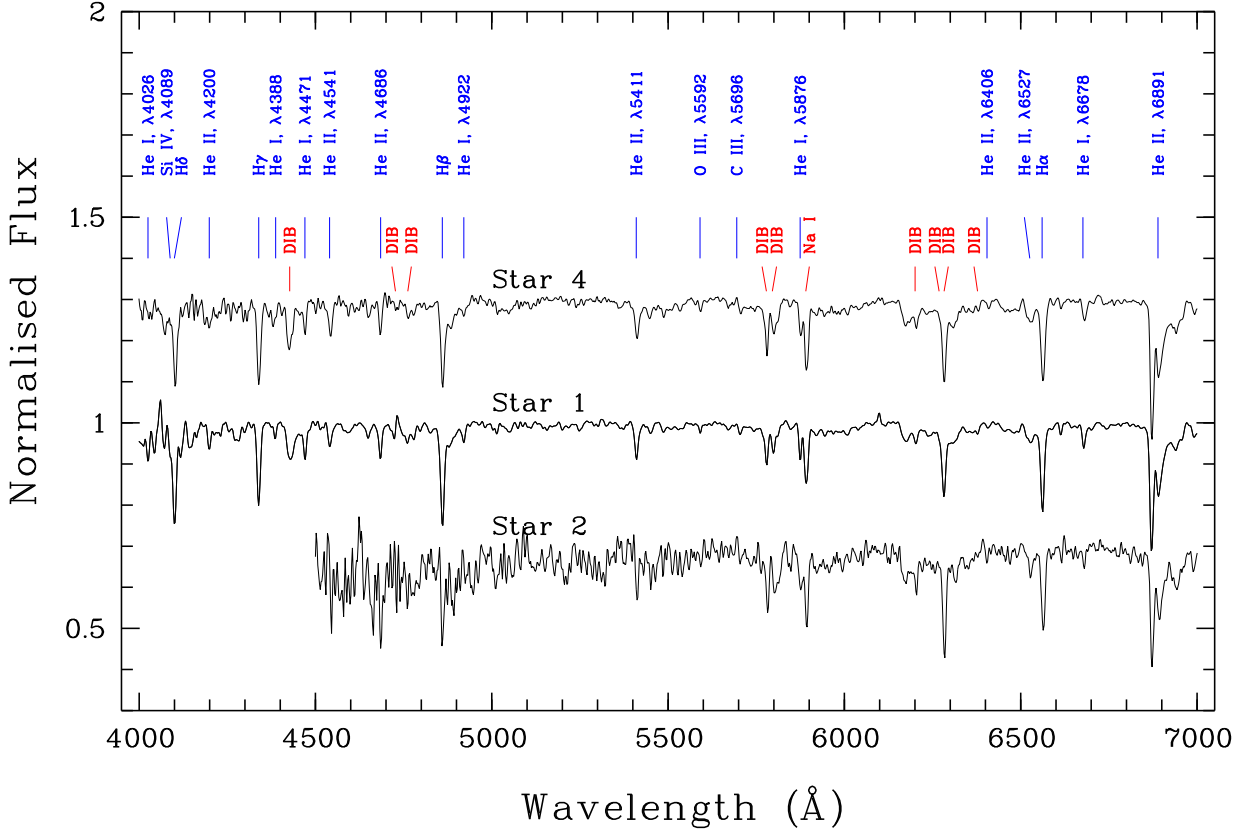}
\caption{Normalized spectra of stars associated with bow shocks\,4
(top), 1 (middle) and 2 (bottom) with principal lines and most
prominent DIBs indicated. The spectra of stars\,4 and 2 are
shifted upwards and downwards by $\simeq 0.3$ continuum flux unit,
respectively.
}
\label{fig:spec}
\end{figure*}

Figure\,\ref{fig:spec} presents the resulting normalized spectra
of stars 1, 2 and 4 in the $\lambda\lambda 4000-7000$ \AA \,
region. All three spectra show numerous narrow absorption lines of
hydrogen, He\,{\sc i} and He\,{\sc ii}, which is typical of O-type
stars. Si\,{\sc iv} $\lambda 4089$ and O\,{\sc iii} $\lambda 5592$
absorptions are also visible in the spectra. We detected the
emission of C\,{\sc iii} $\lambda 5696$ in all three spectra
(which characterizes the stars of the O4--O8 spectral type;
Walborn \cite{wa80}), although in the spectrum of star 2 the
detection is marginal. No forbidden lines can be seen in the
spectra. Numerous diffuse interstellar bands (DIBs) are present in
the spectra. A blend of strong Na\,{\sc i} $\lambda 5890, 96$
absorption lines is of circumstellar or interstellar origin, while
the strong absorptions visible at $\lambda > 6800$ \AA \, are
telluric in origin.

To determine the spectral types of stars\,1 and 4, we used the
traditional classification criteria based on the ratio of the
equivalent widths (EWs) of H\,{\sc i} $\lambda 4471$ and He\,{\sc
ii} $\lambda 4541$ lines (Morgan, Keenan \& Kellman \cite{mo43};
Conti \& Alschuler \cite{co71}). Using Table\,3 in Conti \&
Alschuler (\cite{co71}) and EWs given in Table\,\ref{tab:ew}, we
found the spectral types of stars\,1 and 4 to be O7.5 and O6.5,
respectively, with an uncertainty of $\pm 1$ subtype. The more
earlier subtype inferred for star\,4 is consistent with the
increase of the C\,{\sc iii} $\lambda 5696$ emission line strength
with luminosity (Walborn \cite{wa80}). Similarly, using Table\,5
in Conti \& Alschuler (\cite{co71}) and EWs of Si\,{\sc iv}
$\lambda 4089$ and He\,{\sc i} $\lambda 4143$ lines from
Table\,\ref{tab:ew}, we found that both stars are supergiants.
This inference, however, is inconclusive because of the large
errors in the measurements of the Si\,{\sc iv} $\lambda 4089$
line, which allow the two stars to belong to any luminosity class
(see also below).

The poor observational conditions and the relatively high
extinction towards star\,2 (see Table\,\ref{tab:phot}) degraded
the spectrum of the star, which precludes us from using
traditional classification criteria. Instead, we utilized the
classification scheme for O stars in the yellow-green
($\lambda\lambda 4800-5420$ \AA) proposed by Kerton, Ballantyne \&
Martin (\cite{ke99}), which is based on the ratio of EWs of the
lines He\,{\sc i} $\lambda 4922$ and He\,{\sc ii} $\lambda 5411$:
\begin{equation}
{\rm SpT}=(9.04\pm 0.10)+(4.10\pm 0.23)\log{\rm EW_{4922}}/{\rm
EW_{5411}} \, . \label{eqn:1}
\end{equation}
Using Eq.\,(\ref{eqn:1}) and EWs given in Table\,\ref{tab:ew}, we
found SpT$\simeq 5.5$. Similarly, applying this equation for
stars\,1 and 4 gives their spectral types of O8 and O7,
respectively, which agree well with the estimates based on the
traditional classification criteria. The spectral type can also be
estimated using separately EW$_{4922}$ and EW$_{5411}$ (Kerton et
al. \cite{ke99}):
\begin{equation}
{\rm SpT}=(4.82\pm 0.03)+(7.92\pm 0.56){\rm EW_{4922}} \, ,
\label{eqn:2}
\end{equation}
\begin{equation}
{\rm SpT}=(12.80\pm 0.41)-(7.08\pm 0.58){\rm EW_{5411}} \, .
\label{eqn:3}
\end{equation}
From Eqs.\,(\ref{eqn:2}) and (\ref{eqn:3}), we found SpT$\simeq
6.0$ and 5.0, respectively; both estimates are consistent with
that derived from Eq.\,(\ref{eqn:1}). Still, although it is clear
that star\,2 is of O-type, the estimates of its spectral type
suffer from the large errors of measured EWs (see
Table\,\ref{tab:ew}).

\begin{table*}
\caption{Equivalent widths (EWs; in \AA) of absorption lines used
for spectral classification of bow-shock-producing stars 1, 2 and
4.} \label{tab:ew}
\begin{center}
\begin{minipage}{\textwidth}
\begin{tabular}{ccccccc}
\hline \hline
Star & Si\,{\sc iv} $\lambda 4089$ & He\,{\sc i} $\lambda 4143$ &
He\,{\sc i} $\lambda 4471$ & He\,{\sc ii} $\lambda 4541$ &
He\,{\sc i} $\lambda 4922$ & He\,{\sc ii} $\lambda 5411$ \\
\hline
1 & 0.62$\pm$0.35 & 0.30$\pm$0.15 & 0.90$\pm$0.12 & 0.75$\pm$0.09 & 0.58$\pm$0.08 & 1.07$\pm$0.05 \\
2 & -- & -- & -- & -- & 0.15$\pm$0.14 & 1.07$\pm$0.25 \\
4 & 0.53$\pm$0.34 & 0.14$\pm$0.06 & 0.70$\pm$0.12 & 1.00$\pm$0.12 & 0.34$\pm$0.13 & 1.15$\pm$0.05  \\
\hline
\end{tabular}
\end{minipage}
\end{center}
\end{table*}

The lack of an exact spectral classification does not allow us to
compare the spectroscopic distances to stars\,1, 2 and 4 with the
distance to NGC\,6357. Instead, we use the assumption that these
stars were ejected from NGC\,6357 to constrain their spectral
types and luminosity classes. Using Eq.\,(\ref{eqn:AK}) and
adopting the intrinsic colour, for O stars, $(J-K_{\rm s})_0
=-0.21$ mag (Martins \& Plez \cite{ma06}), we calculated the
$K_{\rm s}$-band extinction, $A_{K_{\rm s}}$, towards these stars
and their absolute magnitudes $M_{K_{\rm s}}$ (see
Table\,\ref{tab:phot}). Then using the calibration of absolute
infrared magnitudes for Galactic O stars by Martins \& Plez
(\cite{ma06}), we found that all three stars should be on the
main-sequence with spectral types of O7\,V, O6.5--7\,V, and O6\,V,
which agree well with those derived from spectroscopy (see
Table\,\ref{tab:phot}, where the photometric spectral types are
enclosed in brackets).

\subsection{Constraining the spectral types of the other bow-shock-producing stars}
\label{sec:spect}

Similarly, assuming that the other bow-shock-producing stars
around NGC\,6357 are located at the same distance as this \hii
region, we calculated $A_{K_{\rm s}}$ and $M_{K_{\rm s}}$ for
these stars and estimated their photometric spectral types (see
Table\,{\ref{tab:phot}). For stars\,3 and 5 we found spectral
types of O5\,V and O6.5\,V, respectively. The former estimate
agrees reasonably well with the spectroscopically derived spectral
type, while the latter one is inconsistent with the spectral
classification reported by Neckel (\cite{ne84}). We note, however,
that the spectral types given by Neckel (\cite{ne84}) are
systematically later than those derived from the more recent
observations (Massey et al. \cite{ma01}; Damke et al.
\cite{da06}). For example, one of the brightest (and most massive)
stars in the cluster AH03\,J1725$-$34.4, [N78]\,51, was classified
by Neckel (\cite{ne84}) as an O9 star, while in the recent study
of this cluster by Damke et al. (\cite{da06}) the spectral type of
the star was found to be O3.5\,V((f)).

Applying the same procedure for stars\,6 and 7 suggests that both
of them could be of the same spectral type of B0\,V (so that the
smaller size of the bow shocks generated by stars\,6 and 7 could
be because of weaker winds of these stars). And finally, we
estimated the spectral type of the candidate bow-shock-producing
star (star\,8) to be O9--9.5\,V. Additional spectroscopic and
photometric observations of the bow-shock-producing stars around
NGC\,6357 are warranted.

\subsection{Proper motions and likely parent clusters of bow-shock-producing stars}
\label{sec:prop}

\begin{table*}
\caption{Proper-motion measurements for the seven
bow-shock-producing stars (stars\,1--7) and one candidate
bow-shock-producing star (star\,8) around NGC\,6357. Two
measurements are given for each star to indicate the uncertainties
in the measurements. For each data set, the peculiar (transverse)
velocities (in Galactic coordinates) were calculated and added to
the table. A likely parent cluster for each star is indicated in
the last column.} \label{tab:cat}
\renewcommand{\footnoterule}{}  
\begin{tabular}{lcccccccc}
\hline \hline Star &  $\mu _\alpha \cos \delta$ & $\mu _\delta$ &
Ref. & $v_{\rm l}$ & $v_{\rm b}$ & Parent cluster \\
~ & mas ${\rm yr}^{-1}$ & mas ${\rm yr}^{-1}$ &  & $\kms$ & $\kms$
&  \\ \hline 1 & $-1.5\pm 1.9$  & $-1.6\pm 2.0$   & 1 & $-4.4\pm
16.1 $ & $9.9\pm 15.3$ & AH03\,J1725$-$34.4 \\
1 & $-3.8\pm 3.0$  & $-5.0\pm 3.0$   & 2 & $-7.1\pm 24.2$ & $10.0\pm 24.2$ &  \\
2 & $-8.5\pm 4.7$  & $-3.2\pm 4.7$   & 3 & $-46.8\pm 37.9$ &
$48.5\pm 37.9$ & Pismis\,24 \\
2 & $-8.9\pm 12.2$ & $-11.3\pm 12.2$ & 2 & $-102.4\pm 98.3$ & $14.1\pm 98.3$ & \\
3 (HD\,319881) & $0.0\pm 1.4$ & $-2.8\pm 1.0$ & 1 & $-5.7\pm 8.9$
& $-5.3\pm 10.5$ & Pismis\,24 \\
3 (HD\,319881) & $3.4\pm 2.6$ & $-3.2\pm 2.7$ & 2 & $6.8\pm 21.8$ & $-29.9\pm 21.0$ & & \\
4 & $-4.4\pm 2.0$ & $0.9\pm 1.1$     & 1 & $-1.0\pm 11.8$ &
$39.7\pm 10.4$ & Pismis\,24/AH03\,J1725$-$34.4 \\
4 & $-9.0\pm 2.9$ & $2.3\pm 3.0$     & 2 & $-13.2\pm 23.9$ & $76.5\pm 23.6$ & \\
5 ($[$N78$]$\,34) & $-4.9\pm 4.8$ & $11.7\pm 4.8$    & 3 &
$69.0\pm 38.7$ & $93.1\pm 38.7$ & AH03\,J1725$-$34.4 \\
5 ($[$N78$]$\,34) & $-11.8\pm 12.2$ & $18.7\pm 12.2$ & 2 & $83.9\pm 98.3$ & $170.9\pm 98.3$ & \\
6 & $-4.6\pm 3.8$ & $0.8\pm 2.1$     & 1 & $-2.5\pm 22.3$ &
$41.0\pm 27.0$ & AH03\,J1725$-$34.4 \\
6 & $-8.1\pm 12.2$ & $0.1\pm 12.2$   & 2 & $-23.1\pm 98.3$ & $61.1\pm 98.3$ & \\
7 & $-3.9 \pm 2.1$ & $0.0 \pm 1.9$   & 1 & $-4.7\pm 15.8$ &
$33.1\pm 16.4$ & Pismis\,24 \\
7 & $-7.3 \pm 3.0$ & $-3.2 \pm 3.0$  & 2 & $-41.5\pm 24.2$ & $41.5\pm 24.2$ & \\
\hline 8 & $-6.0\pm 1.8$ & $1.9\pm 2.6$     & 1 & $ -1.4\pm 19.1$
& $ 55.5\pm 16.8$ & Pismis\,24 \\
8 & $-9.4\pm 3.0$ & $-2.5\pm 3.0$    & 2 & $ -46.2\pm 24.2$ & $ 58.1\pm 24.2$ & \\
\hline
\end{tabular}
\tablebib{(1)~UCAC3 (Zacharias et al. \cite{za10}); (2)~PPMXL
(R\"{o}ser et al. \cite{roe10}); (3)~NOMAD (Zacharias et al.
\cite{za04}).}
\end{table*}

In Table\,\ref{tab:cat} we provide two representative proper
motion measurements for each bow-shock-producing star, found with
VizieR. For most stars we used the measurements from the most
recent catalogues, namely UCAC3 (Zacharias et al. \cite{za10}) and
PPMXL (R\"oser, Demleitner \& Schilbach \cite{roe10}). When the
UCAC3 proper motion is marked ``with doubts" (because it relies on
``less than two good matches"), we used the data from the NOMAD
catalogue (Zacharias et al. \cite{za04}). Although most
measurements are insignificant (i.e. the measurement uncertainties
are comparable to the measurements themselves), it is of interest
to check whether or not the orientation of the stellar peculiar
velocities implied by these measurements is consistent with the
distribution of the stars on the sky and with the orientation of
their associated bow shocks.

To convert the observed proper motions into the transverse
peculiar velocities of the stars, we used the Galactic constants
$R_0 = 8.0$ kpc and $\Theta _0 =240 \, {\rm km} \, {\rm s}^{-1}$
(Reid et al. \cite{re09}) and the solar peculiar motion
$(U_{\odot} , V_{\odot} , W_{\odot})=(10.0, 11.0, 7.2) \, {\rm km}
\, {\rm s}^{-1}$ (McMillan \& Binney \cite{mc10}). The derived
velocity components in Galactic coordinates are given in columns 5
and 6 of Table\,\ref{tab:cat}. For the error calculation, only the
errors of the proper motion measurements were considered.
Obviously, the orientation of peculiar velocities of stars\,1, 2,
3, 4, 6 and 8 are consistent (within their margins of error) with
the orientation of their bow shocks and with the possibility that
these stars are moving away from NGC\,6357. For stars\,5 and 7 we
found that their peculiar velocities have "wrong" orientations,
i.e. are inconsistent with the orientation of the associated bow
shocks.

Because of the low significance of proper motion measurements, we
rely only on the orientation of the bow shocks in the
identification of parent clusters for bow-shock-producing stars.
We caution, however, that the close proximity of the clusters on
the sky and the absence of proper motion measurements for the
clusters make the identification ambiguous. Thus, the likely
parent clusters indicated in the last column of
Table\,\ref{tab:cat} should be considered as tentative.

\section{Three circular shells}
\label{sec:shells}

As we mentioned in Sect.\,\ref{sec:search}, one of the by-products
of the search for bow shocks around NGC\,6357 with {\it Spitzer}
is the discovery of three circular shells. The positions of these
shells on the sky are marked in Fig.\,\ref{fig:Pismis24} by
crosses. The appearance of all three shells (see
Figs.\,\ref{fig:LBV-1}--\,\ref{fig:LBV-3}) is typical of
circumstellar shells created by luminous blue variable (LBV) stars
and Wolf-Rayet stars of late WN-types (WNL) (see Gvaramadze et al.
\cite{gv10c} for {\it Spitzer} images of such shells).

\begin{figure}
 \resizebox{9cm}{!}{\includegraphics{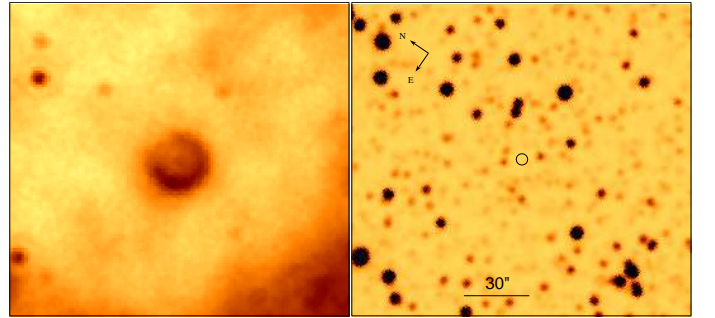}}
 \caption{{\it Left}: MIPS $24\,\mu$m image of shell\,1. {\it Right}: 2MASS $K_{\rm s}$ band image of the
 same field with the position of the geometric centre of shell\,1 indicated by a circle.}
  \label{fig:LBV-1}
\end{figure}
\begin{figure}
 \resizebox{9cm}{!}{\includegraphics{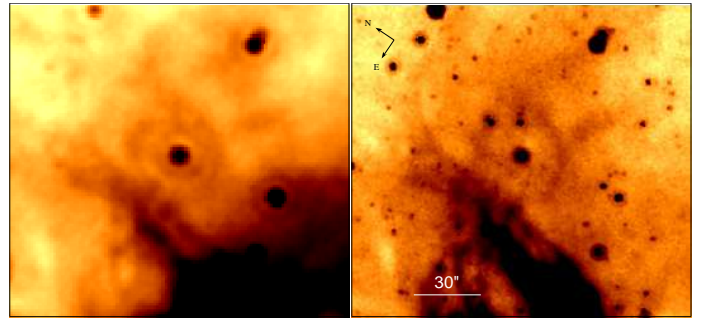}}
 \caption{{\it Left}: MIPS $24\,\mu$m image of shell\,2. {\it Right}: IRAC $8\,\mu$m
 image of the same field.
 }
  \label{fig:LBV-2}
\end{figure}
\begin{figure*}
 \resizebox{18cm}{!}{\includegraphics{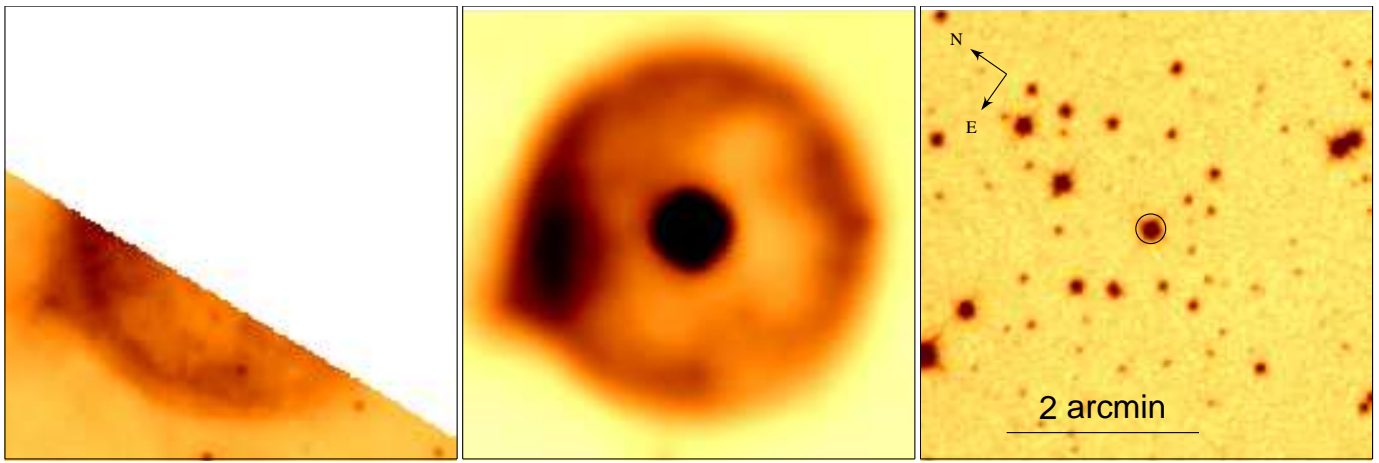}}
 \caption{{\it Left} and {\it Middle}: MIPS $24\,\mu$m and WISE $22\,\mu$m images of shell\,3,
 respectively. The MIPS image is truncated because the MIPSGAL survey covers only a part of the shell.
 {\it Right}: DSS-II (red band) image of the same field. The central star of shell\,3 is marked by a circle.
 }
  \label{fig:LBV-3}
\end{figure*}
\begin{table*}
  \caption{Circular shells around NGC\,6357 and their central stars.}
  \label{tab:LBV}
  \renewcommand{\footnoterule}{}
\begin{tabular}{lccccccccc}
\hline \hline
Shell & RA (2000) & Dec. (2000) & $V$ & $[3.6]$ & $[8.0]$ & Diameter (arcsec) \\
\hline
1 & 17 22 36.7 & $-$33 49 10 & -- & -- & -- & 25 \\
2 & 17 26 19.11$^{a}$ & $-$35 32 37.7$^{a}$ & -- & 13.67$^{a}$ & 8.29$^{a}$ & 40 \\
3 & 17 20 31.76$^{b}$ & $-$33 09 48.9$^{b}$ & 14.3$^{b}$ & -- & -- & 210 \\
\hline
\end{tabular}
 \tablefoot{
 \tablefoottext{a}{GLIMPSE Source Catalog (http://irsa.ipac.caltech.edu/applications/Gator/).}
 \tablefoottext{b}{NOMAD (Zacharias et al. \cite{za04}).}
 }
\end{table*}

One of the shells (hereafter shell\,1) is located at only $\simeq
0\fdg6$ (or $\simeq 18$ pc in projection) to the northwest of
NGC\,6357. The MIPS 24\,$\mu$m image of this shell (see left panel
of Fig.\,\ref{fig:LBV-1}) shows a clear ring-like structure with a
diameter of $\simeq 25$ arcsec and enhanced brightness along the
southeast rim faced towards NGC\,6357. Fig.\,\ref{fig:LBV-1} also
shows the central point-like source being smeared-out somewhat
because of the 6 arcsec resolution of the MIPS 24\,$\mu$m image.
We interpret this source as a massive evolved star physically
related to the shell. Many dozens of similar evolved stars were
recently revealed through the detection of their infrared
circumstellar shells with {\it Spitzer} and follow-up spectroscopy
(Gvaramadze et al. \cite{gv09b}, \cite{gv10a},b,c; Wachter et al.
\cite{wa10}, \cite{wa11}). The DSS-II and 2MASS images, however,
do not show any counterpart to the putative central star of
shell\,1. The most plausible explanation of the absence of the
2MASS counterpart is that the central star is highly absorbed by
the foreground dusty material. In this connection, we note that
the central stars of some circumstellar shells presented in
Gvaramadze et al. (\cite{gv10c}) also cannot be seen in the 2MASS
images, while they are visible in the IRAC (3.6, 4.5, 5.8 and 8.0
$\mu$m) ones. Unfortunately, shell\,1 was not covered by the
GLIMPSE survey, so that at the moment we cannot prove the
existence of the central source that is apparently visible at
$24\,\mu$m. Deep infrared imaging of shell\,1 is therefore highly
desirable. Detection of the central star and its follow-up
spectroscopy would allow us to unveil the nature of the shell and
potentially link the star to the star-forming region NGC\,6357.

If one assumes that shell\,1 is produced by a massive star ejected
from NGC\,6357, then the linear size of the shell is $\simeq 0.2$
pc, which is comparable to that of some LBV shells (e.g. Buemi et
al. \cite{bu10}). If, however, the actual size of the shell is 1-2
pc (which is more typical of circumstellar shells associated with
LBV and WNL stars), then the distance to this shell is $\simeq
7-14$ kpc. At this distance the star would be at a height of
$\simeq 180-360$ pc above the Galactic plane, which would imply
that it is a runaway.

For the limiting $K_{\rm s}$ magnitude of the 2MASS survey of 14.8
and the $K_{\rm s}$-band absolute magnitude of LBV and WNL stars
of $\simeq -9$ and $-6$ mag, respectively (Blum, Depoy \& Sellgren
\cite{bl95}; Crowther et al. \cite{cr06}), one finds that
$A_{K_{\rm s}} \ga 10-13$ mag (or $A_V \ga 90$-110 mag) if the
star is located at 1.7 kpc. An extinction this high could be
understood if the star were located within (or behind) the dense
molecular cloud associated with NGC\,6357 (see
Sect.\,\ref{sec:pismis-24}).

The second shell (hereafter shell\,2) is located at $\simeq
1\fdg3$ to the south of NGC\,6357. Fig.\,\ref{fig:LBV-2} shows the
MIPS 24\,$\mu$m (right panel) and IRAC 8\,$\mu$m images of this
shell. The ring-like structure of shell\,2 is more obvious in the
8\,$\mu$m image, from which we measured a diameter of the shell of
$\simeq 40$ arcsec ($\simeq 0.3$ pc at a distance of 1.7 kpc). The
central star of shell\,2 is visible not only at 24\,$\mu$m, but
also in all four IRAC bands. We also found a weak counterpart to
this star in the 2MASS $K_{\rm s}$-band image, which, however, is
not listed in the 2MASS All-Sky Catalog of Point Sources
(Skrutskie et al. \cite{sk06}).

The third shell (hereafter shell\,3) is located at $\simeq 1\fdg4$
to the northwest of NGC\,6357. Unfortunately, the MIPSGAL survey
covers only a part of this quite extended ($\simeq 3.5$ arcmin in
diameter) shell with a prominent blow-up to the northeast (see
left panel of Fig.\,\ref{fig:LBV-3}). The full extent of shell\,3,
however, is visible in the WISE 22\,$\mu$m image (middle panel of
Fig.\,\ref{fig:LBV-3}), which shows an almost circular
limb-brightened nebula and its central star (visible also in the
DSS-II image; right panel of Fig.\,\ref{fig:LBV-3}). Using the
SIMBAD database, we identified the central star with the
emission-line star \object{Hen\,3-1383}, classified in the
literature as an Me (The \cite{th61}; Sanduleak \& Stephenson
\cite{sa73}; Allen \cite{al78}) or a candidate symbiotic (Henize
\cite{he76}) star. Our recent optical spectroscopy of Hen\,3-1383
with the Southern African Large Telescope (SALT), however, showed
that the spectrum of this star is almost identical to those of the
prototype LBV \object{P\,Cygni} (see, e.g., Stahl et al.
\cite{st93}) and the recently discovered candidate LBV MN\,112
(Gvaramadze et al. \cite{gv10b}), which implies the LBV
classification for Hen\,3-1383 as well (Gvaramadze et al., in
preparation).

The details of the three shells are summarized in
Table\,\ref{tab:LBV}. In this table we give approximate
coordinates of the geometric centre of shell\,1 and the
coordinates of the central stars associated with two other shells.
For the central star of shell\,2 we give its IRAC 3.6 and 8.0
$\mu$m magnitudes (taken from the GLIMPSE Source
Catalog\footnote{Available at
http://irsa.ipac.caltech.edu/applications/Gator/.}), while for the
central star of shell\,3 we give its visual magnitude from the
NOMAD catalogue. The last column of the table contains the angular
size of the shells.

\section{Discussion}
\label{sec:dis}

We explored the idea that young star clusters lose a significant
fraction of their massive stars because of gravitational few-body
interactions between the cluster members. The efficiency of
dynamical star ejection depends mainly on the star number density
in the cluster core, which is maximum either already during
cluster formation or becomes maximum later on owing to the
Spitzer's mass segregation instability (Spitzer \cite{sp69}). Mass
segregation could be a rapid ($\la 0.5$ Myr) process if the
cluster was born in a compact configuration (e.g. Kroupa
\cite{kr08}), so that dynamical star ejections could be effective
well before the cluster starts to expel its members via
binary-supernova explosions (the latter process becomes efficient
several Myr after the formation of the first massive binary
stars).

The runaway OB stars spread all around the parent cluster
constitute the population of massive field stars. It is believed
that most, if not all, field OB stars were formed in the clustered
way (Lada \& Lada \cite{la03}; Smith, Longmore \& Bonnell
\cite{sm09}) and subsequently found themselves in the field either
because of the dynamical processes in the parent clusters or
because of rapid cluster dissolution (e.g., Kroupa \& Boily
\cite{kr02}; de Wit et al. \cite{de05}; Schilbach \& R\"{o}ser
\cite{sc08}; Gvaramadze \& Bomans \cite{gv08b}; Pflamm-Altenburg
\& Kroupa \cite{pf10}). Detection of massive stars in the field
around young clusters and linking them to these clusters is
therefore important not only for the problem of the origin of
field OB stars, but also for the problem of massive star-formation
in general.

The onset of OB star ejection from a young cluster very much
depends on whether the massive stars form as tight binaries near
the cluster centre or not (Clarke \& Pringle \cite{cl92}). Thus
N-body experiments can be performed with different assumptions on
the OB stellar population. To see which of the theoretical
(numerical) models of cluster dynamical evolution better
correspond to observations, one should identify as many stars
ejected from a given star cluster as possible. The detected
runaway stars would constitute a sample for the future
high-precise proper motion and parallax measurements with the
space astrometry mission {\it Gaia}, which are required to
determine the timing of ejections (e.g. Paper\,I) and thereby to
constrain the initial conditions for young star clusters (Kroupa
\cite{kr08}) and to distinguish between the primordial and the
dynamical origin of mass segregation in these clusters.

In the absence of reliable proper motion measurements for distant
and/or highly obscured field OB stars, the only viable possibility
to prove their runaway status and to identify their parent
clusters is through the detection of bow shocks that are possibly
generated by these stars\footnote{The runaway status of some field
OB stars can also be confirmed via measurement of their high
peculiar radial velocities, but this channel for detection of
runaways does not allow us to determine the star's direction of
motion on the sky and therefore cannot be used to identify their
birth clusters.}. The characteristic scale of bow shocks is
determined (for the given number density of the ambient
interstellar medium and the peculiar velocity of the star) by the
strength of the stellar wind, which makes the massive stars prone
for producing {\it detectable} bow shocks. On the other hand, the
fact that dynamical star ejection is most efficient in
mass-segregated clusters (Oh et al., in preparation) implies that
during the early dynamical evolution of these clusters the ejected
stars are preferentially the most massive ones. This inference is
consistent with the observational fact that the percentage of O
stars among the runaways is higher than that of B stars (e.g.
Stone \cite{st91}) and could also be derived from the existence of
an extended (tens of parsecs) halo of very massive (O2-type) stars
around the very young ($\sim 1$-2 Myr) star cluster R136 in the
Large Magellanic Cloud (e.g. Brandl et al. \cite{br07}; Walborn et
al. \cite{wa02}).

Our search for OB stars running away from the star-forming region
NGC\,6357 has resulted in the discovery of seven bow shocks. The
morphology of the bow shocks suggests that the associated stars
were ejected from clusters Pismis\,24 and AH03\,J1725$-$34.4,
which are embedded in NGC\,6357. The actual number of ejected
stars could be much higher because only a small fraction of
runaway stars produce observable bow shocks (see
Sect.\,\ref{sec:search}). Four of five (or even all; see
Sect.\,\ref{sec:spect}) bow-shock-producing stars with
spectroscopically determined spectral types turn out to be O-type
stars, which provides additional evidence that young massive star
clusters expel preferentially the most massive stars.

Follow-up spectroscopy of three bow-shock-producing stars detected
around NGC\,6357 showed that all of them are O-type stars. Thus,
the search for bow shocks around young clusters and follow-up
spectroscopy of their associated stars is a useful tool for
detection of highly obscured, yet unknown OB stars. In some cases,
the bow-shock-producing stars are not necessarily high-velocity
runaways. For example, high-velocity outflows driven from young
star clusters by the collective effect of stellar winds and
ionizing emission can create bow shocks around (low-velocity; not
necessarily massive) stars in the cluster's halo. In this case,
the bow shocks are facing towards the cluster's centre. Numerous
examples of such bow shocks were detected around young clusters
and OB associations: \object{Orion Cluster} (Bally, O'Dell \&
McCaughrean \cite{ba00}), \object{Trumpler\,14} (Ascenso et al.
\cite{as07}), \object{M17} and \object{RCW\,49} (Povich et al.
\cite{po08}), \object{Cyg\,OB2} (Kobulnicky, Gilbert \& Kiminki
\cite{ko10}). One can also imagine the situation when a slowly
($\sim 10 \, \kms$) moving massive star encounters a dense medium
(e.g. the remainder of the parent molecular cloud) and its wind
starts to interact with the ionized gas outflowing from the cloud
surface towards the star with a velocity comparable to the sound
speed ($\simeq 10 \, \kms$). In the reference frame of this gas,
the stellar motion is supersonic so that a bow shock is formed
ahead of the (slowly) moving star. We speculate that bow shock\,1
was probably generated in this way. This possibility is suggested
by the small separation ($\simeq 10$ pc in projection) of star\,1
from its likely birth cluster AH03\,J1725$-$34.4, which implies
that the peculiar velocity of the star could be as low as $5-10 \,
\kms$, provided that it was ejected from the cluster $\sim 1-2$
Myr ago (cf. Table\,\ref{tab:cat}). Moreover, the right panel of
Fig.\,\ref{fig:bow1} shows that bow shock\,1 and its associated
star are located at the periphery of the \hii region, from which
one can infer that star\,1 has met and ionized a density
enhancement on its way.

\begin{figure}
 \resizebox{9cm}{!}{\includegraphics{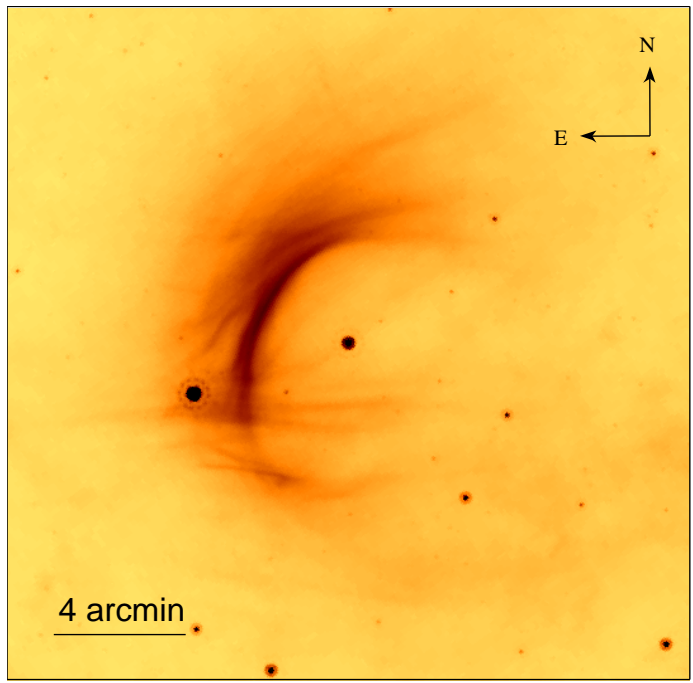}}
 \caption{MIPS $24\,\mu$m image of the bow shock generated by the BC0.7\,Ia star
 HD\,2905 and cirrus-like filaments around it (see text for details).
 }
  \label{fig:HD2905}
\end{figure}

Finally, we discuss the origin of the gap at the leading edge of
bow shock\,4 and the curious cirrus-like filaments around it. In
Gvaramadze et al. (\cite{gv11b}), we suggested that similar
cirrus-like filaments around the bow shock generated by the
high-mass X-ray binary \object{Vela\,X-1} (see Fig.\,2 in
Gvaramadze et al. \cite{gv11b}) are caused by interstellar dust
grains aligned with the local interstellar magnetic field and
heated by the radiation of Vela\,X-1. This suggestion implies that
bow shock\,4 propagates along the local interstellar magnetic
field, which affects the compression ratio of the shocked gas by
making it less dense along the flanks of the bow shock (e.g.
Draine \& McKee \cite{dr93}). The density asymmetry in turn could
affect the heating of the dust grains accumulated behind the front
of bow shock\,4 and thereby could be responsible for its
brightness asymmetry. In the absence of reliable data on
parameters of bow shock\,4 and its associated star we refrain from
a more detailed discussion of the problem. We note, however, that
similar gaps are typical of many other bow shocks. One of them
(although less impressive and not at the symmetry axis of the bow
shock) is shown in Fig.\,\ref{fig:HD2905}, which presents the MIPS
$24\,\mu$m image (Program Id.: 30088, PI: A.Noriega-Crespo) of a
bow shock produced by the BC0.7\,Ia (Walborn 1972) star
\object{HD\,2905}. Like bow shock\,4, the bow shock around
HD\,2905 is surrounded by cirrus-like filaments, some of which are
apparently attached to the front of the bow shock near the gap. To
conclude, if our suggestion on the origin of gaps is correct, then
the cirrus-like filaments around bow shocks could serve as tracers
of the local magnetic field, while the bow-shock-producing stars
could serve as probes of the Galactic magnetic field.

\section{Summary}
\label{sec:sum}

Search for bow-shock-producing stars around the star-forming
region NGC\,6357 with two embedded young ($\sim 1-2$ Myr) star
clusters Pismis\,24 and AH03\,J1725$-$34.4 resulted in the
discovery of seven bow shocks. The geometry of the bow shocks
strongly suggests that all seven associated stars were ejected
from NGC\,6357. Because only a small fraction of runaway stars
produce detectable bow shocks, the actual number of massive stars
ejected from NGC\,6357 could several times exceed the number of
detected bow-shock-producing stars. This inference is consistent
with the theoretical expectation that young star clusters lose a
significant fraction of their massive star content at the very
beginning of their dynamical evolution. Follow-up spectroscopy of
three detected bow-shock-producing stars showed that all of them
are O-type stars. This result is consistent with the observational
fact that O stars prevail among the runaways and implies that
search for bow shocks around massive clusters provides a useful
tool for discovery of new OB stars. A by-product of our search for
bow shocks is the detection of three circular shells around
NGC\,6357, whose morphology is typical of LBV and WNL stars.
Follow-up spectroscopy of the central star of one of these shells
showed that its spectrum is almost identical to that of the
prototype LBV P\,Cygni, which implies the LBV classification for
this star as well.

\begin{acknowledgements}

We are grateful to the referee for comments that allowed us to
improve the presentation of the paper. VVG acknowledges financial
support from the Deutsche Forschungsgemeinschaft. AYK acknowledges
support from the National Research Foundation of South Africa.
This research has made use of the NASA/IPAC Infrared Science
Archive, which is operated by the Jet Propulsion Laboratory,
California Institute of Technology, under contract with the
National Aeronautics and Space Administration, the SIMBAD database
and the VizieR catalogue access tool, both operated at CDS,
Strasbourg, France.
\end{acknowledgements}

\end{document}